\documentclass[3p,preprint,authoryear]{elsarticle}

\usepackage{amssymb}
\usepackage{graphicx}
\usepackage{caption}
\usepackage{subcaption}

\newcommand*{\addheight}[2][.5ex]{%
  \raisebox{0pt}[\dimexpr\height+(#1)\relax]{#2}%
}

\biboptions{comma}

\usepackage[figuresright]{rotating}

\begin{document}
\pagestyle{empty}

\begin{frontmatter}



 \title{Random walk model from the point of view of algorithmic trading}

\author{Oleh Danyliv}
\author{Bruce Bland}
\author{Alexandre Argenson}
\address{Fidessa group plc, One Old Jewry, London, EC2R 8DN,
United Kingdom} \tnotetext[t1]{Authors are grateful to Xavier
Cochi at Fidessa group plc for valuable discussions and initiation
of this project. Authors thank Mathew Lynch, our intern from
University of Bath, for a technical assistance.}
\tnotetext[t2] {The views expressed in this article are those of the authors and do not necessarily reflect the views
of Fidessa group plc, or any of its subsidiaries.}

\begin{abstract}
Despite the fact that an intraday market price distribution is not
normal, the random walk model of price behaviour is as important
for the understanding of basic principles of the market as the
pendulum model is a starting point of many fundamental theories in
physics. This model is a good zero order approximation for liquid
fast moving markets where the queue position is less important
than the price action. In this paper we present an exact solution
for the cost of the static passive slice execution. It is shown,
that if a price has a random walk behaviour, there is no optimal
limit level for an order execution: all levels have the same
execution cost as an immediate aggressive execution at the
beginning of the slice. Additionally the estimations for the risk
of a limit order as well as the probability of a limit order
execution as functions of the slice time and standard deviation of
the price are derived.
\end{abstract}


\end{frontmatter}

\pagestyle{headings} \setcounter{page}{1}

\section{Introduction}

A standard strategy for impact avoiding algorithms like TWAP or
VWAP (\cite{johnson2010}) is to split the order size into smaller
child orders (slices) which are traded using execution strategies
(see, for example, \cite{markov2012}). A slice is therefore an
order with a small volume and a small market impact which tries to
capture the bid-ask spread where possible.  The {\it simple
passive strategy} for a child order is the passive post and wait
strategy, where on the initial phase the order is traded passively
at a specific level and if not filled passively it is filled
aggressively at the end of the slice interval. Aggressive
execution at the end of the time interval is associated with the
penalty, which the trader has to pay because of the price moving
away from the limit price of the order.

\cite{jeria2009} describe Goldman Sachs'  Piccolo algorithm which
is a good example of a passive execution where is no order queue:
it's passive leg creates buy below ask and sell above bid limit
orders which are then executed aggressively if not filled. Their
analysis of 19,821 passive and 6,919 aggressive child orders
showed that the passive execution decreases market impact, but in
fact, is not better than an immediate aggressive execution: all in
cost of passive orders was estimated at 4.6 bps and the all-in
cost of aggressive orders was 4.1 bps. Although 48\% of orders
captured spread, the clean-up cost of non-filled orders was high.

This document models mathematically the basic process of passive
execution where the price is assumed to follow a symmetrical
random walk, and uses this mathematical modelling to analyse the
effects of limit prices on the variance of costs. It should be
noted that the approach taken ignores the queue position of orders
arriving in order books,  however  is still a good proxy for fast
moving liquid markets on which the majority of trading in the
worlds markets is carried out.

\section{Probabilities on the binary tree}

\begin{figure}[htp]
    \centering
    \includegraphics[width=0.31\textwidth]{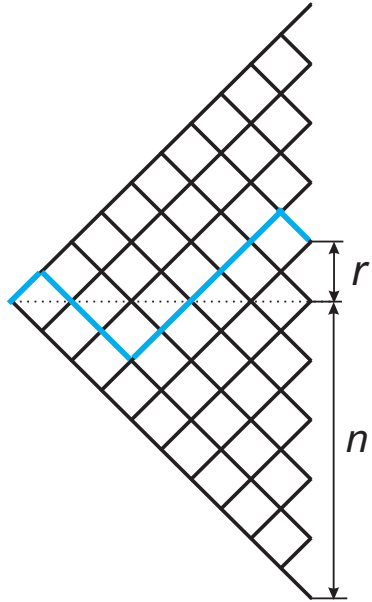}
        \caption{The security price modelled as a walk along a binary tree: it starts at zero and
finishes at value $r$ after $n$ steps.}
    \label{figure1}
\end{figure}

Let us assume that the probability of the price to move up or down
is the same and is $\frac{1}{2}$. The size of the binary tree
shown on Fig. \ref{figure1} is $n$ meaning that the price should
make $n$ random steps in total and final price $r$ is distributed
in a range from $-n$ to $n$.

During the $n$-step long random walk process, the price would make
$n_{\uparrow}$ steps up and $n_{\downarrow}$ steps down. If the
final price is $r$, the following equation should be satisfied:
\begin{equation}
n_{\uparrow}- n_{\downarrow} = r \label{n_up_minus_n_down}\,.
\end {equation}
The total number of steps is $n$, giving the second equaiton
\begin{equation}
n_{\uparrow} + n_{\downarrow} = n \label{n_up_plus_n_down}\,.
\end{equation}
From (\ref{n_up_minus_n_down}) and (\ref{n_up_plus_n_down}) it is
easy to obtain that $n_{\uparrow}=  (n+r)/2$.

The random walk along a binary tree could be constructed as a
random choice of up or down moves on every step (equivalent to
coin tosses). To reach final value $r$, the price has to make
$n_{\uparrow}$ up moves from possible $n$ (fixed amount of "heads"
out of $n$ coin tosses in the fair coin flip analogy). This number
is described by the binomial coefficient

\[
C_n^{n_{\uparrow}}=\frac{n!}{n_{\uparrow} ! (n-n_{\uparrow} )!}\,.
\]
The total number of possible steps on the binary tree is $2^n$.
Therefore, the probability of the price having the final value $r$
making $n$ random steps (as shown on Fig.1) is
\begin{equation}
P_n (r)=\frac{1}{2^n} C_n^{n_{\uparrow}} = \frac{1}{2^n}
C_n^{\frac{n+r}{2}}\,.
 \label{probability}
\end{equation}

\begin{figure}[htp]
    \centering

\begin{tabular}{c c}
      \addheight{\includegraphics[width=0.4\textwidth]{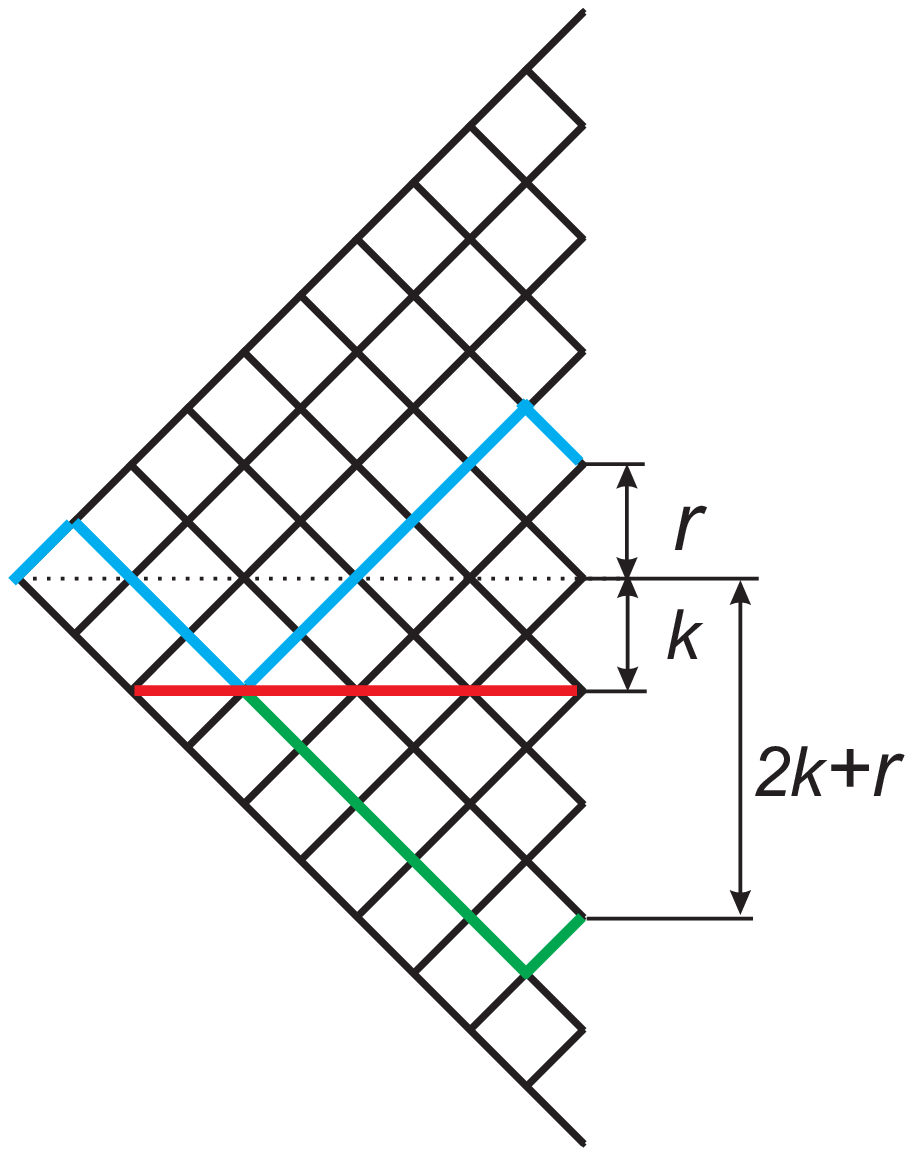}} &
      \addheight{\includegraphics[width=0.4\textwidth]{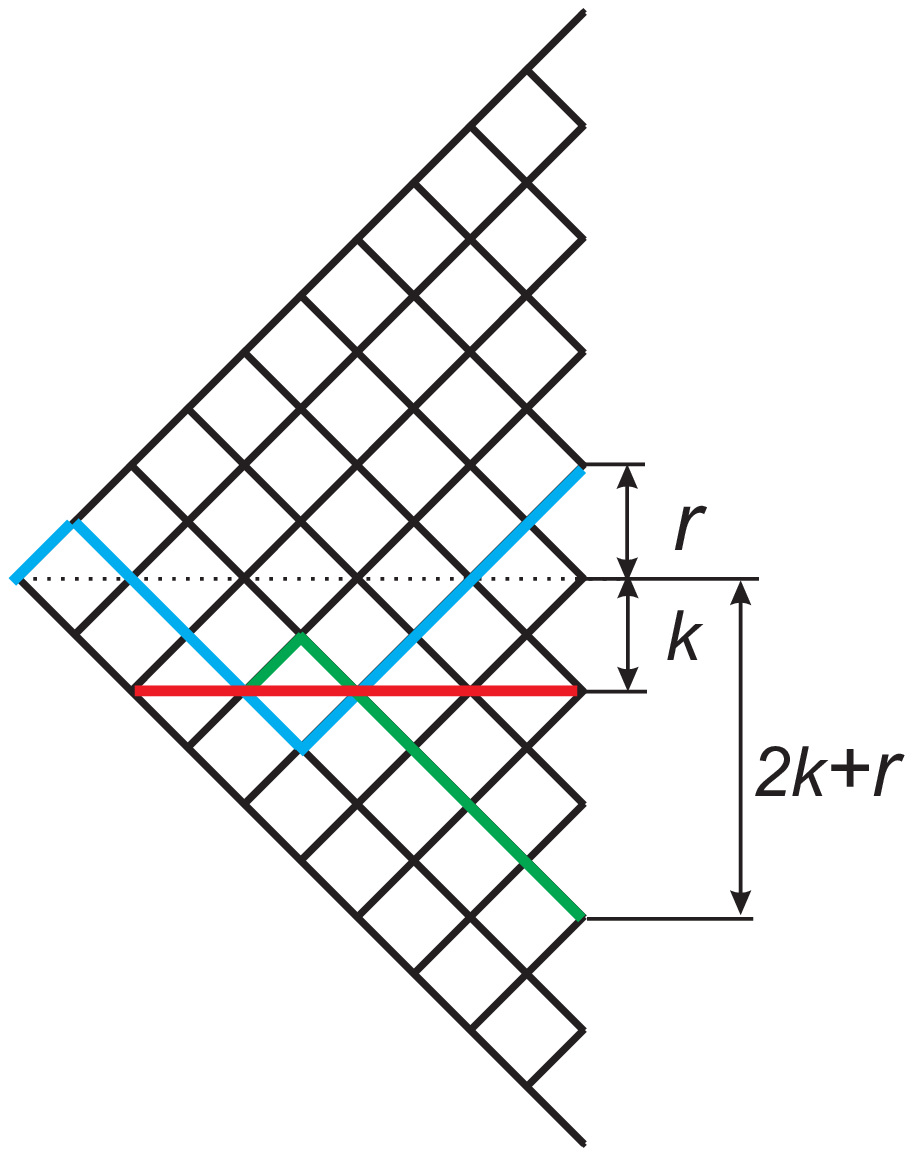}}
\end{tabular}
        \caption{Price could either touch (left diagram) or penetrate
(right diagram) the limit level $k$. The blue lines show direct
trajectories, the green lines show reflected trajectories which
all have the same probabilities of realization.}
    \label{figure2}
\end{figure}

This formula is valid for positive and negative $r$ ($r$ is
negative when the price finishes below the starting level). To
evaluate  the price of limit executions, the probability of the
price to have value $r$ after touching (or penetrating) level $k$
is required. The blue line on the left diagram on
Fig.\ref{figure2} illustrates the price touching limit level $k$
and finishing at value $r$, the right diagram shows the price
penetrating level $k$.

This probability could be calculated using the technique called
{\it reflection} which is based on the fact that the probabilities
of direct and reflected paths are the same (\cite{feller1959}).
When the price reaches level $k$, it has a choice to move up or
down with the same probability. The probability of the real blue
path and the probability of reflected path shown by the green line
on Fig. \ref{figure2} are equivalent. The green trajectory
finishes at point $2k+r$. Using formula (\ref{probability}), the
probability of a price trajectory to touch level $k$ and then
finish at a level $r$ is
\begin{equation}
P_n (r | k)=P_n (2k+r)=\frac{1}{2^n}  C_n^{\frac{n+2k+r}{2}}\,.
\label{prob_touch}
\end{equation}

To calculate the probability of the price to have the final value
$r$ without reaching level $k$, one has to count all the
trajectories which finish at $r$ and subtract the number of
trajectories which finish in $r$ but touched or crossed limit
level $k$. In terms of probabilities, the final result can be
obtained using subtraction (\ref{prob_touch}) from
(\ref{probability}):
\begin{equation}
P_n (r | \overline{k})=P_n (r)-P_n (r | k) = \frac{1}{2^n}
\left(C_n^{\frac{n+r}{2}} - C_n^{\frac{n+2k+r}{2}} \right)\,.
\label{prob_no_touch}
\end{equation}
The same formula  will be valid for cases when $r$ is negative. If
$r$ is a big positive number, there will be cases when no
trajectories which end at $r$ could possibly touch or cross level
$k$. Fig. 3 illustrates the limit case when the touching is still
possible. The maximum amount of down moves in this case is
$n_{\downarrow}=k$: the price starts moving down immediately after
the start of the trajectory and then bounces. The number of
positive steps then is $n_{\uparrow}=n-n_{\downarrow}=n-k$. The
price ends at this critical value $r^*$, meaning
$n_{\uparrow}-n_{\downarrow}=r^*$ and  $r^*=n-2k$. For all $r$
larger than this critical value, the probability to reach this
level (without touching $k$) is simply the probability to reach
level $r$
\begin{equation}
P_n (r | \overline{k}) = P_n (r) = \frac{1}{2^n}
C_n^{\frac{n+r}{2}} {\quad \rm if \quad}  r>n-2k\,.
\label{prob_touch_p}
\end{equation}

The probabilities, which were described above, are visualised on
Fig. \ref{figure4}: few limit orders with different limit  prices
are represented by histograms. Sharp peaks on histograms
correspond to the position of the limit order. The $k=1$ case is
when the limit order is one tick away from the opposite side.
According to the chart, almost  70\% of all orders in this case
will be executed passively. The probabilities of not touching the
limit level $P_n (r|\overline{k})$  are smaller and behave
similarly to the distribution of the price of the underlying
instrument (dashed line).

\begin{figure}[htp]
    \centering
    \includegraphics[width=0.31\textwidth]{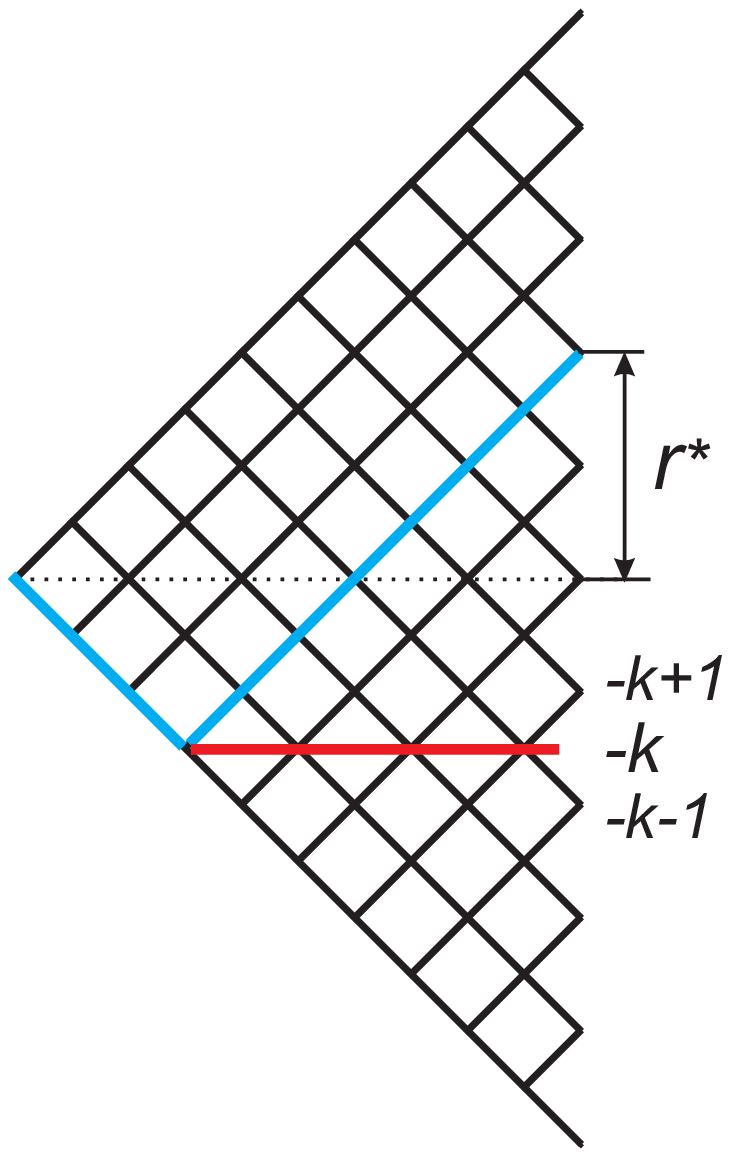}
        \caption{Critical value for $r$. For all $r>r^*$ reaching the touch
level is not possible.}
    \label{figure3}
\end{figure}

\begin{figure}[htp]
    \centering
    \includegraphics[width=0.8 \textwidth]{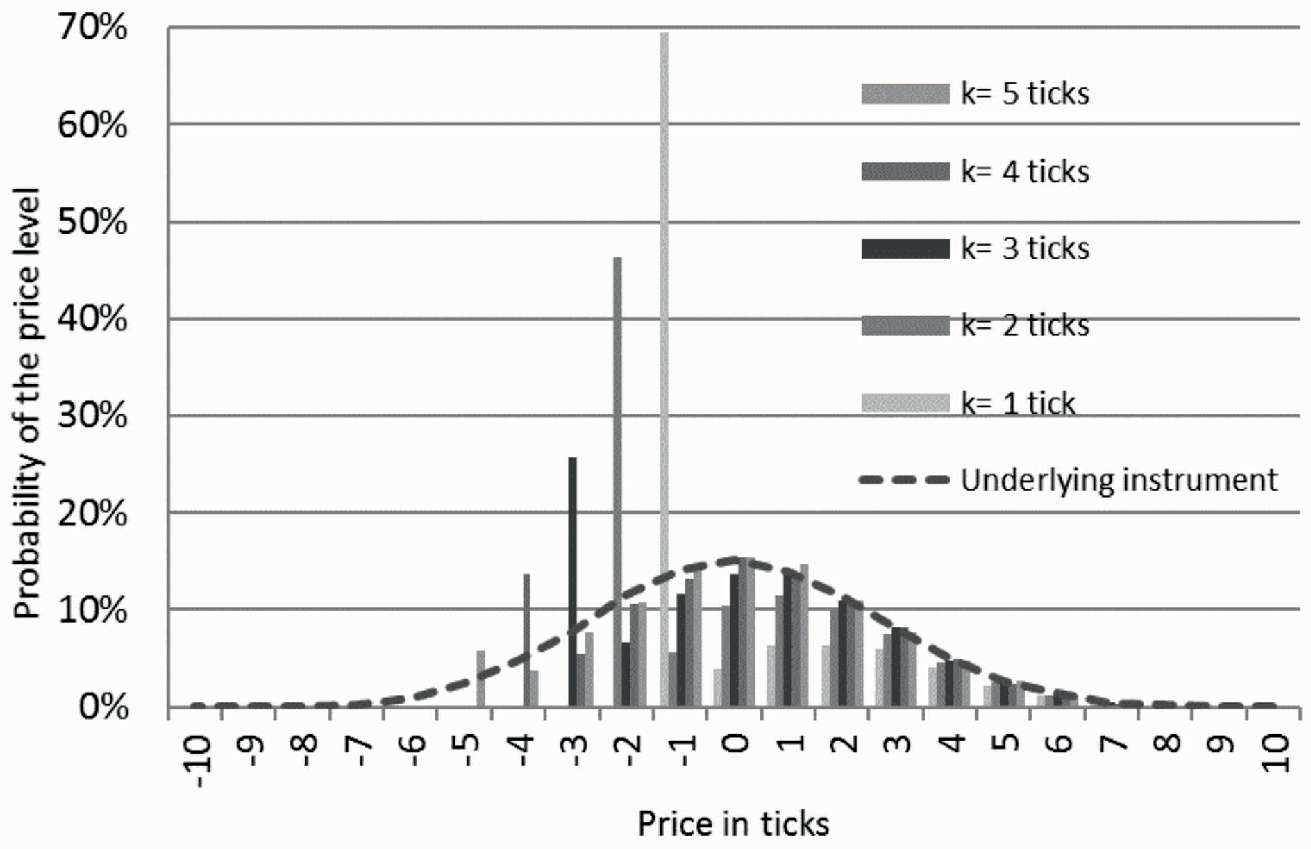}
        \caption{The distribution of the limit order execution prices for the binary tree with $n=10$.}
    \label{figure4}
\end{figure}

\section{The cost of a static limit order execution}

For practical calculations it is useful to shift the starting
price to be zero at the beginning of the random walk. Then,
positive purchasing price would result in a penalty and negative
purchasing price would result in a profit.

Each time the price does not reach a passive level $k$  during the
slice time, the trader will need to go aggressive. The aggressive
price is going to be $r$, since the price ends up on level $r$.
The total penalty for not touching or crossing passive level will
be a sum of all penalties over all possible outcomes $r$. The
final state cannot be smaller than (or equal to) $-k$ and cannot
be larger than the number of steps $n$.

Using (\ref{prob_touch}) and (\ref{prob_touch_p}), the average
cost of execution when the limit level was not reached will be
equal to

\[
\Delta(\overline{k})=\frac{1}{2^n}  \left\{ \sum_{r=-k+1}^{n-2k} r
\left( C_n^{\frac{n+r}{2}} - C_n^{\frac{n+2k+r}{2}}\right) +
\sum_{r=n-2k+1}^n r C_n^{\frac{n+r}{2}} \right\}\,,
\]
or, after regrouping,
\begin{equation}
\Delta(\overline{k})=\frac{1}{2^n}  \left\{ \sum_{r=-k+1}^{n} r
C_n^{\frac{n+r}{2}}  -  \sum_{r=-k+1}^{n-2k} r
C_n^{\frac{n+2k+r}{2}} \right\}\,.
\label{delta_k_dash}
\end{equation}

Every time the price touches (or penetrates) the limit order
price, the order execution price is $-k$. There are two different
possibilities of this situation:

\begin{enumerate}

\item {\bf The price finishes on level $-k$ or below this level.}
In this case all the trajectories result in limit executions and
the probability of this case can be calculated using
(\ref{probability}). The average price of the execution is
$-k\times \frac{1}{2^n}  C_n^{\frac{n+r}{2}}$.

\item {\bf The price touches or penetrates level $-k$, but ends up
above level $-k$.} The probability of this situation is calculated
using formula (\ref{prob_touch}) and the average price of this
type of executions is $-k\times \frac{1}{2^n}
C_n^{\frac{n+2k+r}{2}}$.
\end{enumerate}

The summation over all the possible outcomes of the price action
gives the following result:
\begin{equation}
\Delta (k)=\frac{-k}{2^n}  \left\{  \sum_{r=-n}^{-k}
C_n^{\frac{n+r}{2}}
                            + \sum_{r=-k+1}^{n-2k} C_n^{\frac{n+2k+r}{2}} \right\}\,.
\label{delta_k}
\end{equation}

The total cost of a limit execution consists of the profit from
the situations when the limit price was hit, minus the cost of all
the aggressive orders (trajectories without a touch of the limit
level). Since the price is counted from  zero level, the total
execution cost is equal to the price with a minus sign: limit
executions give negative price, which means profit. Changing the
sign will create a situation where positive values mean a profit:
\begin{equation}
\Delta_k=-(\Delta(\overline{k}))+ \Delta(k)),
\end{equation}
where $\Delta(\overline{k})$ and $\Delta(k)$ are given by
(\ref{delta_k_dash}) and (\ref{delta_k}).

Using mathematical induction it is possible to prove that
$\Delta_k=0$, in other words, that the benefit from passive
execution is exactly the same as the loss from situations when the
limit order was not touched.

By using variables substitution $r\rightarrow r-2k$ in the
expressions (\ref{delta_k_dash}) and (\ref{delta_k}), the
difference of profits for level $k$ and the next level $k+1$ could
be written as the following:

\begin{equation}
\Delta_{k+1}- \Delta_k  =\frac{1}{2^n}  {\sum_{r=k+1}^n
C_n^{\frac{n+r}{2}} -\sum_{r=-n}^{-k-1} C_n^{\frac{n+r}{2}}}\,.
\end{equation}
This expression is equal to zero since
$C_n^{\frac{n+r}{2}}=C_n^{\frac{n-r}{2}}$. Therefore for all $k$
\begin{equation}
\Delta_{k+1} = \Delta_k\,.
\end{equation}

The situation when $k=0$ corresponds to the scenario where the
order is placed at the immediate aggressive price for which we
know that $\Delta_{k=0}=0$. Consequently, $ \Delta_k=0$ for all
$k$ .

That proves the fact that a passive execution of the slice has no
optimal level: all levels always results in a zero gain. This is
equivalent to the immediate aggressive execution at the beginning
of the slice. This result explains observations of the performance
of the Piccolo trading algorithm (Jeria,  Schouwenaars, Sofianos
[2009]).

\section{The risk of the limit order execution}

The standard way is to consider the standard deviation
$\sigma_X^2$ of trade outcomes $\delta(r,k)$ as a risk measure of
the execution. It should be noted, that $\delta(r,k)$ are not
normally distributed for the simple passive strategy.

\begin{figure}
     \centering
     \begin{subfigure}[b]{0.45\textwidth}
         \centering
         \includegraphics[width=\textwidth]{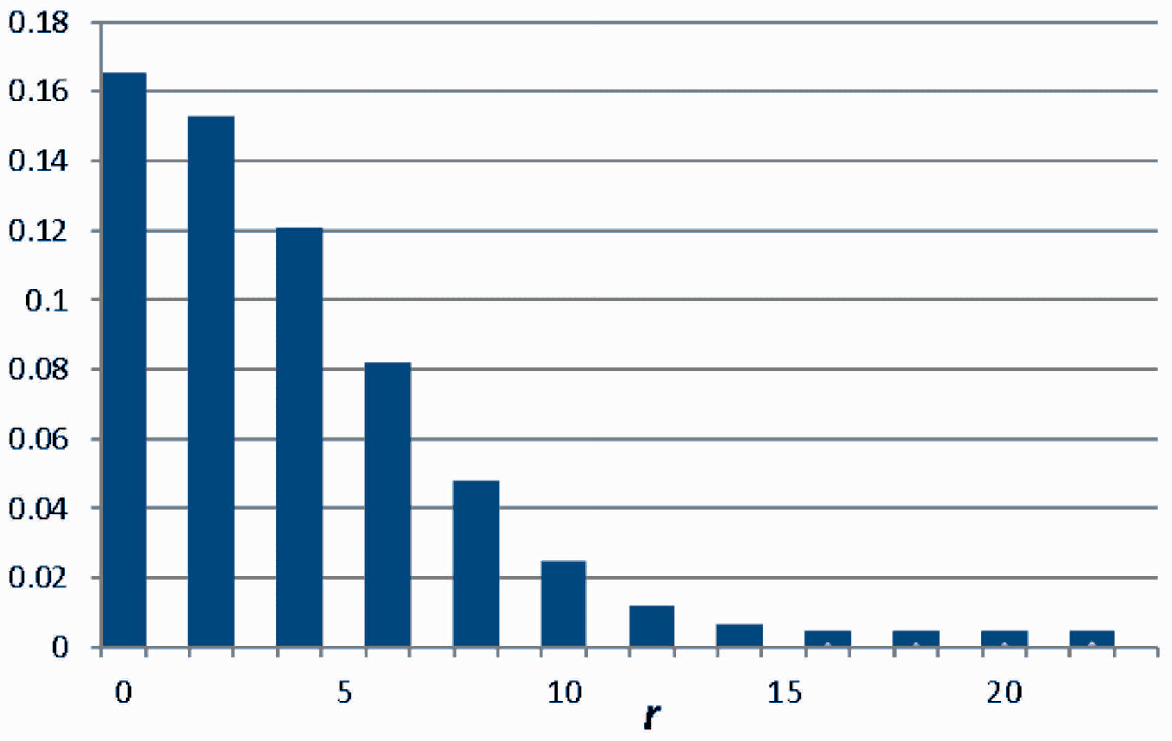}
         \caption{Probabilities $P_n (r)$ for even $n$. The trajectory can finish only in values 0, 2, 4, $\dots$}
         \label{figure5}
     \end{subfigure}
     \hfill
     \begin{subfigure}[b]{0.45\textwidth}
         \centering
         \includegraphics[width=\textwidth]{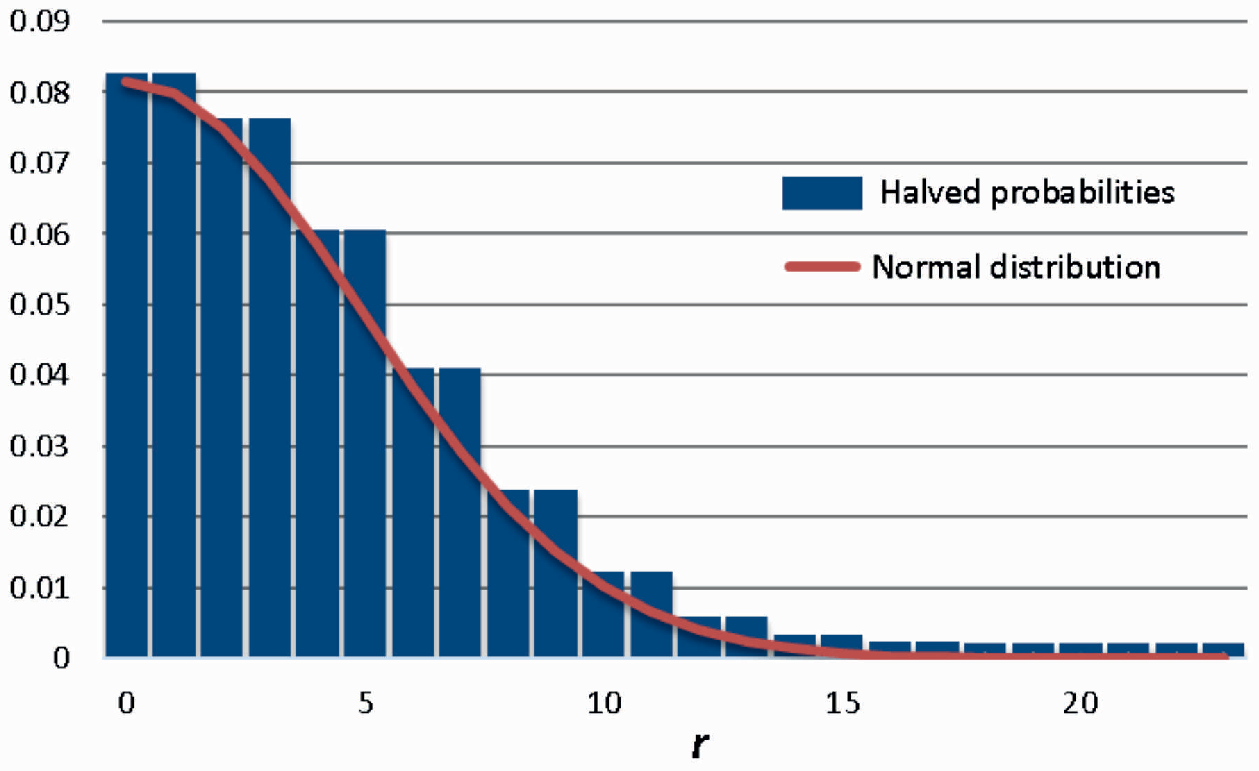}
         \caption{Halved probabilities $P_n (r)$ which are distributed at every point and
corresponding normal distribution  $N(0,\sqrt{n})$.}
         \label{figure6}
     \end{subfigure}
        \caption{Fitting probabilities $P_n (r)$ with normal distribution for binary tree.}
        \label{figure5and6}
\end{figure}

In the previous section we proved that the average outcome over
all possible final values $r$, $\Delta(k) =
\sum_r{\delta(r,k)P_n(r)} = 0$ for all $n$. The variance of the
execution results has a simple form
\begin{equation}
\sigma_X^2= \sum_r (\delta(r,k)-\Delta_{k})^2 P_n(r)=\sum_r
\delta(r,k)^2 P_n(r), \label{variance}
\end{equation}
where the result depends on the length of the binary tree $n$ and
the distance to the limit order $k$. Similarly to the situation
with the average cost of execution, the probability of an outcome
splits into two components (when the limit order was touched and
when the limit level was not reached) $P_n(r)=P_n (r|k)+P_n
(r|\overline{k})$. If the limit order was not hit, then the
outcome of the price run will be equal to the final price $r$. In
analogy with the formula (\ref{delta_k_dash}),
\begin{equation}
\sigma_X^2 (\overline{k}) = \sum_r r^2 P_n (r|\overline{k}) =
\frac{1}{2^n}  \left\{ \sum_{r=-k+1}^n r^2 C_n^{\frac{n+r}{2}} -
\sum_{r=-k+1}^{n-2k} r^2 C_n^{\frac{n+2k+r}{2}} \right \}\,.
\label{sigma_k_}
\end{equation}
If the the order was executed passively, then, using the same
probabilities as in (\ref{delta_k}),
\begin{equation}
\sigma_X^2 (k) = \sum_r (-k)^2 P_n (r|{k}) = \frac{k^2}{2^n}
\left\{ \sum_{r=-n}^{-k}  C_n^{\frac{n+r}{2}} +
\sum_{r=-k+1}^{n-2k} C_n^{\frac{n+2k+r}{2}} \right \}\,.
\label{sigma_k}
\end{equation}
Comparing this formula and formula (\ref{delta_k}) for the average
execution price when the limit level is hit, it is easy to notice
that $\sigma^2 (k)=-k\Delta(k)$. Using proven relation
$\Delta(k)+\Delta(\overline{k})=$0,  $\sigma^2 (k)=
-k\Delta(\overline{k})$, the expression (\ref{sigma_k}) could be
rewritten via sums present in (\ref{sigma_k_}):
\begin{equation}
\sigma_X^2  = \sigma_X^2 (\overline{k})+ \sigma_X^2
(k)=\frac{1}{2^n} \left\{ \sum_{r=-k+1}^n
(r^2+kr)C_n^{\frac{n+r}{2}} - \sum_{r=-k+1}^{n-2k} (r^2+kr)
C_n^{\frac{n+2k+r}{2}} \right\}\,.
\label{sigma1}
\end{equation}
This expression is exact and is valid for even and odd
combinations of parameters $n$ and $k$.
After substitution $r' \rightarrow r + 2k$ in the second term of
(\ref{sigma1}) and simple, but laborious transformations, the
variance could be rewritten in the form which reveals its explicit
dependency on the limit level $k$:

\begin{equation}
\sigma_X^2  = \frac{1}{2^n}  \left\{ 4k \sum_{r=k+1}^n
rC_n^{\frac{n+r}{2}} - 2k^2 \sum_{r=k+1}^{n}  C_n^{\frac{n+r}{2}}
+ \sum_{r=-k+1}^{k} r^2 C_n^{\frac{n+r}{2}} \right\},
\label{sigma2}
\end{equation}
or, using the definition (\ref{probability}) of probabilities $P_n
(r)$,
\begin{equation}
\sigma_X^2  = 4k \sum_{r=k+1}^n r P_n (r) - 2k^2 \sum_{r=k+1}^{n}
P_n (r) + \sum_{r=-k+1}^{k} r^2 P_n (r). \label{sigma3}
\end{equation}
The sum in (\ref{sigma3}) is performed over all possible values
$r$: if the length of the binary tree is an even number, then the
final price position $r$ could be only an even number (see
Fig.\ref{figure5}). If $n$ is an odd number, then possible
trajectories ending could only be odd numbers. The level of
passive order $k$ is independent and could be any number in the
range $[0,n]$.

The first term in the expression (\ref{sigma3}) provides the main
contribution for small values of $k$ and it is linear by  this
parameter. The second term is smaller and corresponds to the
second order of level $k$. It will be shown that the third term
corresponds to the cube of level $k$.

\section{Analytical approximation for the variance}

\begin{figure}[htp]
    \centering
    \includegraphics[width=0.7 \textwidth]{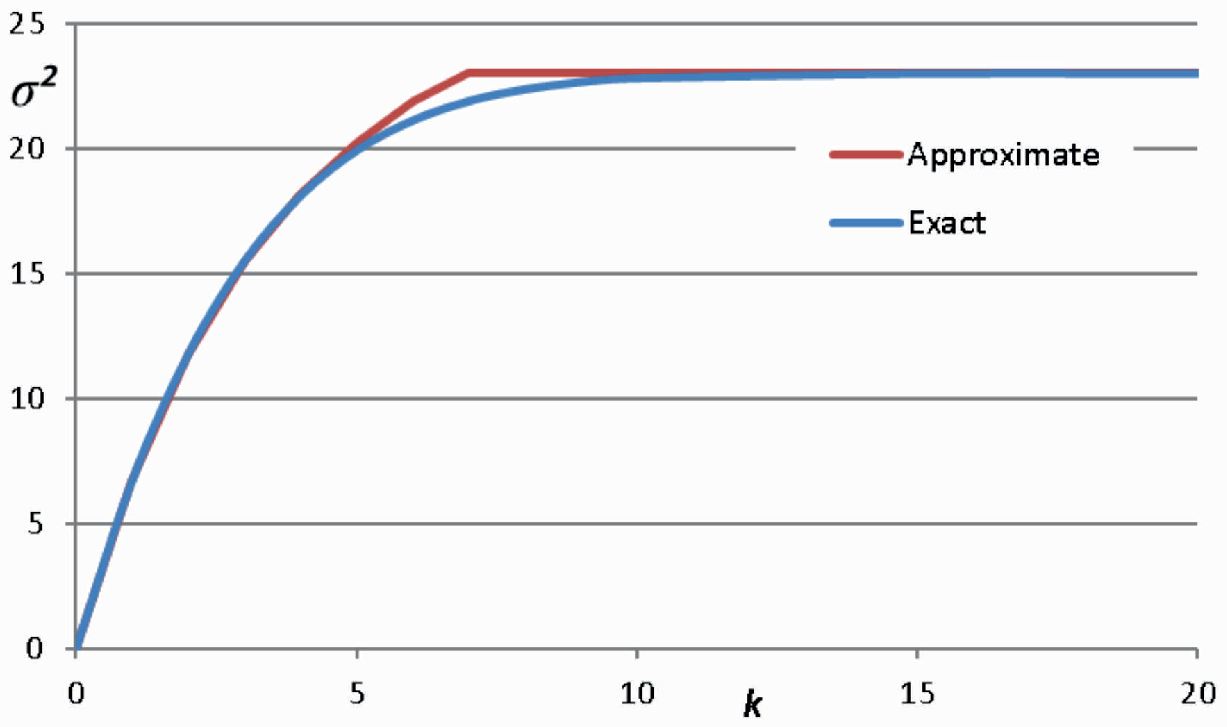}
        \caption{An approximate (\ref{sigma_approx}) and exact (\ref{sigma3})
variances for different limit order levels. They start from zero
when the order is executed aggressively at the beginning of the
slice and reach maximum value 23 for the selected binary tree size
$n=23$. The approximate value is capped by the maximum possible
value 23.}
    \label{figure7}
\end{figure}

The summation in (\ref{sigma3}) which is performed over possible
$r$ could be approximated by a summation over all values $r$ with
halved probability (see Fig.\ref{figure6}).
If the tree is large, then the sums with probabilities could be
further substituted by the integral over normal distribution
$N(0,\sqrt{n})$. In the limit of a large binary tree ($n
\rightarrow \infty $) and small values of limit level ($k
\rightarrow 0$) the sums could be simplified as the following:
\begin{eqnarray}
\sum_{r=k+1}^n r P_n (r) &\rightarrow& \frac{1}{\sqrt{2\pi n}}\int_{k}^{\infty} r e^{-\frac{r^2}{2n}}dr = \sqrt{\frac{n}{2\pi}} + O\left(\frac{k^2}{n}\right), \nonumber \\
\sum_{r=k+1}^n P_n (r) &\rightarrow& \frac{1}{\sqrt{2\pi n}}\int_{k}^{\infty} e^{-\frac{r^2}{2n}}dr = \frac{1}{2} + O\left(\frac{k}{\sqrt{n}}\right) \nonumber, \\
\sum_{r=-k+1}^{k} r^2 P_n (r) &\rightarrow& \frac{1}{\sqrt{2\pi
n}}\int_{-k}^{k} r^2 e^{-\frac{r^2}{2n}}dr  = k^2\times
O\left(\frac{k}{\sqrt{n}}\right). \nonumber
\end{eqnarray}
Substituting the values of approximated integrals into expression
(\ref{sigma3}) and terms up to second order over $k$, the
approximate formula for standard deviation of the limit order at
level $k$ is
\begin{equation}
\sigma_X^2 \approx 4k\sqrt{\frac{n}{2\pi}} - k^2\,.
\label{sigma_approx}
\end{equation}
This approximation should be capped with the maximum possible
value $n$, which corresponds to the case where level $k \ge n$.
Then the limit level is never reached and  the variance of the
limit order is equal to the variance of the underlying price
$\sigma^2= n$. Adding this limitation to the approximation
(\ref{sigma_approx}) will make it work in the whole range of
values $k$ as it is shown on Fig.\ref{figure7}.

The first term in (\ref{sigma_approx}), which is linear over limit
level $k$,  is providing the main contribution. Therefore, the
variance of the limit order results $\sigma_X^2 \propto k$  for
levels close to the touch, which are the most important during
algo trading.  Additionally, in real systems a random walk of
length $n$ will last the order time $T$. This time describes the
dynamic of the system and we will call it a {\it sample time}. The
standard deviation of the price during the order time is going to
be
\begin{equation}
\sigma(T) = \sqrt{n}.
\label{sigma_T}
\end{equation}
Using this relationship, one can link the standard deviation of
the execution $\sigma_X(T)$ during the order time $T$ with the
standard deviation of price via
\begin{equation}
\sigma_X(T) \approx \sqrt{\frac{4k\sigma(T)}{\sqrt{2\pi}}} ,
\label{sigma_approx1}
\end{equation}
where the distance of the limit order to opposite side of the
market $k$ is measured in absolute price units.

From another point of view, the size of the system $n$ in real
systems  is proportional to the time of the order $T$ and formula
(\ref{sigma_approx}) shows that even putting the limit order one
tick away from touch creates a risk which can be expressed via
deviation of results as $\sigma(k=1)  \propto T^{\frac{1}{4}}$.
That fact that even touch orders could possibly have large
standard deviation advocates for using dynamic order placing
(pegging and adaptive strategies).

\section{Probability of a passive fill}
The average price of passive executions $\Delta (k)$ calculated in
(\ref{delta_k}) is the product of the price of limit order ($-k$)
multiplied by probability of the passive fill $P(k)$. Therefore
\begin{equation}
 P (k)=\frac{1}{2^n}  \left\{  \sum_{r=-n}^{-k} C_n^{\frac{n+r}{2}}
                            + \sum_{r=-k+1}^{n-2k} C_n^{\frac{n+2k+r}{2}} \right\}
\label{p_k}
\end{equation}

Both terms in this expressions are almost identical.
Mathematically this can be shown by substituting $r' \rightarrow
-r$ in the first term and $r'\rightarrow r+2k$ in the second term:

\begin{equation}
 P (k)=\frac{1}{2^n}  \left\{  \sum_{r=k}^{n} C_n^{\frac{n+r}{2}}
                            + \sum_{r=k+1}^{n} C_n^{\frac{n+r}{2}} \right\}
\label{p_k2}
\end{equation}

The difference between first and the second term in (\ref{p_k2})
is insignificant. It is equal to zero exactly when $n$ and $k$
have different parity (for example, $n$ is even and $k$ is odd).
That could be seen from Fig.\ref{figure3}: $r$ cannot be equal to
$-k$ with the first available value $r = -k+1$ and $\sum_{r=k} =
\sum_{r=k+1}$. In practical calculations, since parameter $n$ is
large, we can always select $n$ slightly larger to change its
parity. Therefore,
\begin{equation}
 P (k)=\frac{2}{2^n}  \left\{  \sum_{r=k+1}^{n} C_n^{\frac{n+r}{2}}
                             \right\} = 2   \sum_{r=k+1}^{n} P_n(r),
\label{p_k3}
\end{equation}
where $P_n(r)$ is the probability to reach point $r$ at the end of
the random walk. Converting this sum into an integral, using the
procedure described in the previous section (but without the
transition $k\rightarrow 0$),
\begin{equation}
P(k, n) = 2 \sum_{r=k+1}^n P_n (r) \rightarrow \frac{2}{\sqrt{2\pi
n}}\int_{k}^{\infty}  e^{-\frac{r^2}{2n}}dr = 1 - {\rm
erf}\left(\frac{k}{\sqrt{2n}}\right),
\label{probability_n}
\end{equation}
or, using the definition of the standard deviation of the price
(\ref{sigma_T}) during the order time $T$ ,
\begin{equation}
P(k, T) = 1 - {\rm erf}\left(\frac{k}{\sigma(T)\sqrt{2}}\right),
\label{probability_T}
\end{equation}
This result can be used to calculate probability at any time. For
an arbitrary time $t=\tau T$, the new length $n'$ of the binary
tree will be extended/shortened by parameter $\tau$ and
\begin{equation}
\sqrt{n'} = \sqrt{\tau  n} = \sigma(T)\sqrt{\tau}
\label{sigma_tau}
\end{equation}
Substituting this into expression for probability of a passive
fill (\ref{probability_n}), will result in
\begin{equation} P(k,
t=\tau T) =  1 - {\rm
erf}\left(\frac{k}{\sigma(T)\sqrt{2\tau}}\right),
\label{probab_k_t}
\end{equation}
where $\sigma(T)$ is the standard deviation of the price during a
sample time $T$.

One can consider important cases of the result (\ref{probab_k_t})

\begin{enumerate}
\item {
\begin{equation}
P(k, t\rightarrow \infty) =  1.
\end{equation} }
If time of the execution goes to infinity, the price will always
hit the limit level.  This corresponds to the well known fact that
the random walk particle eventually returns to the origin. This
principle, applied to algo trading will read {\it any finite limit
level in random walk model will be executed passively if the time
of the order is infinite}. Unfortunately, this will not happen in
practice because the time of the order is always limited.

\item {
\begin{equation}
P(k=\sigma(T), t=T) =  1 - {\rm
erf}\left(\frac{1}{\sqrt{2}}\right) \approx 32\% ,
\end{equation}
} If the limit order is on the distance of a standard deviation of
the price measured for a sample time $T$, then the probability of
a passive execution during this time is approximately equal to
32\%.

\item {
\begin{equation}
P(k=\sigma(T), t=2 T) =  1 - {\rm erf}\left(\frac{1}{2}\right)
\approx 48\% ,
\end{equation}
} If the limit order is on the distance of a standard deviation of
the price measured for a sample time $T$, then the probability of
a passive execution during the double of this time is
approximately equal to 48\% (roughly half of limit orders will
have a passive fill).
\end{enumerate}

\section{Conclusions}

In this paper we provided an analytical solution to describe the
executed price distribution of a strategy where a limit order is
placed $k$ ticks away from the best opposite market price and
eventually amended at the market price after an elapsed time $T$
if it was not passively filled in between. The analytical solution
assumes the price of the underlying instrument follows a random
walk process (binomial tree).

The analytical solution shows that the average executed price is
always equal to the aggressive market price at the time the
strategy is initiated, regardless of the value of $k$. It also
shows the variance of the executed prices increases with the
distance $k$, as well as with the duration $T$. That corresponds
to growing risk of the execution.

Consequently, the best price point for this strategy is the market
aggressive price at inception. Working a limit price $k$ ticks
away from the initial aggressive price only increases the
dispersion of the results without adding any improvements to the
average executed price. This conclusion is true for very liquid
active markets when the price volatility is much larger than the
tick size and the spread size: for illiquid instruments the effect
of the queue positioning  becomes  as important as price
fluctuations . This effect introduces additional complexity and is
out of the scope of this study.

Therefore, a successful  impact avoiding strategy should be
complemented with a second layer of market data analysis which
proactively decides the best timing to aggress the market  (order
book imbalance, trades acceleration, etc.). This layer should
provide an improved average executed price while trying to
minimize the increased variance of the results.


\bibliographystyle{elsarticle-num}

\end{document}